\documentclass[12pt]{article}

\begin{document}
\begin{center}
{\bf Solutions of Podolsky's Electrodynamics Equation in the First-Order Formalism}\\
\vspace{5mm}
 S. I. Kruglov \\
\vspace{3mm}
\textit{University of Toronto at Scarborough,\\ Physical and Environmental Sciences Department, \\
1265 Military Trail, Toronto, Ontario, Canada M1C 1A4}
\end{center}

\begin{abstract}
The Podolsky generalized electrodynamics with higher derivatives
is formulated in the first-order formalism. The first-order
relativistic wave equation in the 20-dimensional matrix form is
derived. We prove that the matrices of the equation obey the
Petiau-Duffin-Kemmer algebra. The Hermitianizing matrix and
Lagrangian in the first-order formalism are given. The projection
operators extracting solutions of field equations for states with
definite energy-momentum and spin projections are obtained, and we
find the density matrix for the massive state. The $13\times 13$-matrix
Schrodinger form of the equation is derived, and the Hamiltonian is obtained.
Projection operators extracting the physical eigenvalues of the Hamiltonian are found.
\end{abstract}

\section{Introduction}
There is currently a renewal of interest in higher derivative (HD)
field theories. HD field equations appear in many models such as
renormalizable quantum gravity \cite{Stelle}, Podolsky's
generalized electrodynamics \cite{Podolsky}, the Lee-Wick model
\cite{Lee} and others. One of the reasons to consider HD theories
is to improve renormalization properties of theories and to remove
ultraviolet divergences \cite{Thirring}. However, HD models suffer
some difficulties connected with the presence of ghosts. These can
lead to the violation of unitarity \cite{Pais}, \cite{Heisenberg}.
Nevertheless, in some HD models these problems with negative
probabilities and S-matrix unitarity can be avoided
\cite{Hawking}. Also, the quadratic divergence associated with the
Higgs mass were removed in the HD Lee-Wick standard model
\cite{Grinstein}, that solves the hierarchy problem. Extensions of
the minimal standard model living to new physics are justified
until observations at the Large Hadron Collider (LHC) will be
analyzed.

It is well known that in classical electrodynamics the electromagnetic mass
is infinite and, therefore, there are infinities associated with a point particle.
One of the ways to solve this problem in classical theory is to use the Lorentz invariant regularization
of the Maxwell equations at short distances. With the help of an appropriate
cutoff the point particle limit can be achieved. This programme was realized in
Podolsky's electrodynamics. Firstly the interest to Podolsky's electrodynamics was
due to the finiteness of the theory: the electromagnetic energy of a point charge is finite
contrarily to ordinary electrodynamics. If distances are much greater than a cutoff,
Podolsky's electrodynamics converts into Maxwell's electrodynamics.
The solution to Poisson equation for the potential
corresponding to a point charge $e$, located at the origin, in Podolsky's electrostatics is given by \cite{Podolsky}
\[
\varphi=\frac{e}{4\pi r}\left(1-e^{-r/a}\right),
\]
where $a$ is a new parameter of the theory with the dimension of the length playing the role
of the cutoff. This potential becomes the Coulomb potential at distances much bigger than $a$.
At $r\rightarrow 0$ the potential $\varphi$ approaches the finite value $e/4\pi a$.
The energy of the field for a point charge is also finite in the hole space. Thus, the electrostatic
energy can be considered as the regularized electromagnetic mass of a point charge. It was shown \cite{Frenkel}
that higher derivatives terms in Podolsky's equations suppress unphysical runaway solutions with
exponentially growing acceleration of the Abraham-Lorentz equation. There are not unwanted solutions if
the cutoff is greater than half of the electron classical radius. The upper bound on the parameter
is of the order $a\sim 10^{-16}$ cm \cite{Frenkel}, i.e. the same as the Compton wavelength of the neutral $Z$-boson.
Classical Maxwell's electrodynamics is not valid at small distances and time intervals due to quantum effects.
It was also mentioned in \cite{Moniz} that in the framework of non-relativistic quantum theory a natural cutoff
of order of the electron Compton wavelength is effectively appeared by QED processes in close analogy
with the classical theory of extended charges. Thus, one may treat the classical Podolsky's electrodynamics
as an effective theory where a cutoff introduced, $a$, is due to the quantum processes at small distances (large momentum). If distances are larger than $a$, the classical regime begins.

At the same time although QED describes all experimental data well, there are some internal
difficulties with the regularization \cite{Itzykson}. We mention infrared catastrophe: when the average
number of photons $\overline{n}\rightarrow \infty$, then the matrix element $|<0~ out|0~ in> |\rightarrow 0$,
and it is impossible to construct the "out" Fock space from the "in" space, nor to find unitary operator $S$ \cite{Itzykson}. The authors \cite{Itzykson1} wrote: "\textit{There is an alternative possibility to avoid infrared divergences.
We give the photon a small mass $\mu$. This will cut off the low-energy region since now $k^0>\mu$ and therefore remove
the infrared divergence.... The infrared divergences of quantum electrodynamics are essentially classical}".
We continue with the citation \cite{Itzykson2}: "\textit{Another aspect of infrared singularities related
to the long-range character of the Coulomb forces. The latter induces an infinite phase shift on the scattered plane
waves. To prevent it we may introduce a screening factor which in a \textbf{consistent theory} would be related to the
fictitious photon mass $\mu$}". Thus, in QED the cutoff is introduced "by hands" as for small distances
(to remove ultraviolet divergences) as well as for large distances (to avoid infrared catastrophe). Therefore, one may consider naturally to extend classical Podolsky's electrodynamics on the quantum level where the cutoff is appeared due to the presence of higher derivatives. Anyway, different aspects of Podolsky's electrodynamics, in our opinion, have a definite theoretical interest.

Some features of Podolsky's electrodynamics were investigated in \cite{Galvao},
\cite{Barcelos-Neto}, \cite{Accioly}, \cite{Cuzinatto}. The goal of this paper is to formulate
Podolsky's electrodynamics equation in the form of the first-order
relativistic wave equation, and to obtain solutions in the form of
projection matrices.

The paper is organized as follows. In Sec. 2, the third-order field equation
is discussed. We derive the first-order relativistic wave equation for Podolsky's
electrodynamics in the 20-dimensional matrix form. The
Hermitianizing matrix and the Lagrangian in the matrix form are found in
Sec. 3. The projection operators extracting solutions of field
equations for definite energy and spin states of particles are
obtained in Sec. 4. We find the density matrix for the massive
state. In Sec. 5 the $13\times 13$-matrix Schrodinger form of the equation is derived, and
the Hamiltonian is obtained.
Solutions of this equation are found in the form of projection operators.
The results are discussed in Sec. 6. In Appendix A, we consider the first-order wave equation in the presence
of the charge current density. The Lorentz covariance of the equation is proven. Some
useful products of matrices are derived in Appendix B. We obtain ``minimal" polynomials
of the matrix of the equation for massless and massive states. In Appendix C
the ``minimal" polynomial of the Hamiltonian matrix is derived.

The Heaviside's units are chosen, $\hbar =c=1$, and Euclidian metric is used,
$x_\mu=(x_m,ix_0)$. Greek letters range from $1$ to $4$ and Latin
letters range from $1$ to $3$, and there is a summation on
repeated indexes.

\section{Field equations}

\subsection{Third-order field equations}

The Lagrangian of Podolsky's electrodynamics is given by \cite{Podolsky}
\begin{equation}
{\cal
L}_P=-\frac{1}{2}\left[\frac{1}{2}F_{\mu\nu}^2+a^2\left(\partial_\mu
F_{\nu\mu} \right)^2\right], \label{1}
\end{equation}
where $ F_{\mu\nu}=\partial_\mu A_\nu-\partial_\nu A_\mu$ is the
field strength, $\partial_\nu =\partial/\partial x_\nu
=(\partial/\partial x_m,\partial/\partial (it))$. The dimensional
parameter $a$ can be written as $a=1/m$, where $m$ is the mass
parameter. The Euler-Lagrange equations follow from Eq.(1):
\begin{equation}
\left(\partial_\alpha^2 -m^2 \right)\partial_\mu F_{\nu\mu}=0.
\label{2}
\end{equation}
The Lagrangian (1) and the equation of motion (1) are
gage-invariant under the $U(1)$-group.
We can represent Eq.(2), in the momentum space as the matrix equation:
\begin{equation}
\left(p^2+m^2\right)
\left(p^2-p\cdot p\right) A=0,
\label{3}
\end{equation}
were $A=\{A_\mu\}$, the matrix-dyad $p\cdot p$, with matrix elements
$(p\cdot p)_{\mu\nu}=p_\mu p_\nu$, is introduced, and the four-momentum being $p_\mu=(\textbf{p},ip_0)$.
The matrix $M=p^2-p\cdot p$ obeys the minimal polynomial $M(M-p^2)=0$, so that the eigenvalues of the matrix $M$ are zero and $p^2$. Thus, Eq.(3) leads to the dispersion equation:
\begin{equation}
p^2\left( p^2+m^2\right) =0.  \label{4}
\end{equation}
Eq.(4) shows that there are massless and massive states in the spectrum.
The propagator of fields is given by
\begin{equation}
\frac{m^2}{p^2\left( p^2+m^2\right)}=\frac{1}{p^2}-\frac{1}{p^2+m^2}.  \label{5}
\end{equation}
The first term in Eq.(5) is the propagator of the photon massless field
and the second term corresponds to the propagator of the massive
state of the field. A "wrong" sign $(-)$ in Eq.(5)
indicates that the massive field state is a ghost.
As a result, the massive field state gives the negative contribution to the energy \cite{Podolsky},
and the classical Hamiltonian is unbounded. To have the positive eigenvalues of the
Hamiltonian in the second quantized theory, one has to introduce the indefinite metric.
The commutation relations for creation, annihilation operators of the massive state have the wrong sign
$(-)$ \cite{Podolsky}. The Hilbert space of states is the direct sum of
the two subspaces $H_p$ and $H_n$ with positive ($H_p$) and negative
($H_n$) square norms. The massless states correspond
to a positive square norm, and massive states $-$
to a negative square norm. The transitions between two subspaces $H_p$ and $H_n$
break the unitarity of the theory. But if the mass $m\rightarrow \infty$
such transitions are forbidden and the unitarity is recovered.
Thus, a ghost can be removed in the theory at large $m$. This procedure is similar to
the Pauli-Villars regularization of Feynman diagrams. Therefore, there is physical sense
of the Podolsky theory. We also can argue (similar to Lee-Wick model
\cite{Grinstein}) that there is no problem with unitarity if the massive
photon decays to ordinary fermions through its couplings and is not in the spectrum.

\subsection{First-order field equations}

Now we reformulate the third-order field equation (2) in the form of first-order
relativistic wave equation. Let us consider the system of first-order equations
\begin{equation}
\partial_\mu\psi_{\nu\mu} + m \widetilde{\psi}_\nu=0,
\label{6}
\end{equation}
\begin{equation}
\partial_\nu\psi_\mu-\partial_\mu\psi_\nu +m \psi_{\mu\nu}=0,
\label{7}
\end{equation}
\begin{equation}
\partial_\mu\widetilde{\psi}_{\nu\mu} + m \widetilde{\psi}_\nu=0,
\label{8}
\end{equation}
\begin{equation}
\partial_\nu\widetilde{\psi}_\mu-\partial_\mu\widetilde{\psi}_\nu +m
\widetilde{\psi}_{\mu\nu}=0,
\label{9}
\end{equation}
where
\begin{equation}
\psi_\mu=mA_\mu,~~~ \psi_{\mu\nu}=F_{\mu\nu},
~~~\widetilde{\psi}_\mu=\frac{1}{m}\partial_\nu F_{\nu\mu},
~~~\widetilde{\psi}_{\mu\nu}=\frac{1}{m^2}\partial_\alpha^2
F_{\mu\nu}.
 \label{10}
\end{equation}
After replacing $\widetilde{\psi}_\nu$ from Eq.(6) and $\widetilde{\psi}_{\mu\nu}$ from Eq.(9) into Eq.(8), one obtains Eq.(2). Eq.(7) is the usual equation for the potentials. Thus, we claim that the system of first-order equations (6)-(9) is equivalent to the third-order Eq.(2). Let us introduce the
20-dimensional wave function
\begin{equation}\Psi (x)=\left\{ \psi _A(x)\right\} =\left(
\begin{array}{c}
\psi_\mu (x)\\
\psi_{\mu\nu}(x)\\
\widetilde{\psi}_\mu (x)\\
\widetilde{\psi}_{\mu\nu} (x)
\end{array}
\right) \hspace{0.5in}(A=\mu , [\mu\nu] ,\widetilde{\mu}
,\widetilde{[\mu\nu]}) ,\label{11}
\end{equation}
where $\psi_{[\mu\nu]} (x)=\psi_{\mu\nu} (x)$,
$\psi_{\widetilde{\mu}} (x)=\widetilde{\psi}_\mu (x)$,
$\psi_{\widetilde{[\mu\nu]}} (x)=\widetilde{\psi}_{\mu\nu} (x)$.
The function $\Psi(x)$ represents the direct sum of two
four-vectors $\psi_\mu (x)$, $\widetilde{\psi}_\mu(x)$, and two
antisymmetric tensors of the second rank $\psi_{\mu\nu} (x)$,
$\widetilde{\psi}_{\mu\nu} (x)$.

We explore the elements of the entire matrix algebra $\varepsilon
^{A,B}$ \cite{Bogush}, \cite{Kruglov} with matrix elements and
products
\begin{equation}
\left( \varepsilon ^{M,N}\right) _{AB}=\delta _{MA}\delta _{NB},
\hspace{0.5in}\varepsilon ^{M,A}\varepsilon ^{B,N}=\delta
_{AB}\varepsilon ^{M,N}, \label{12}
\end{equation}
where $A,B,M,N=\mu,[\mu\nu],\widetilde{\mu},\widetilde{[\mu\nu]}$,
and generalized Kronecker symbols
\[
\delta_{[\mu\nu][\alpha\beta]}=\delta_{\mu\alpha}\delta_{\nu\beta}- \delta_{\mu\beta}\delta_{\nu\alpha}.
\]
The $\varepsilon ^{M,N}$ are $20\times 20$-matrices, that consist
of zeros and only one element is unity where the row $M$ and the
column $N$ cross.

With the help of Eq.(11),(12) the system of equations (6)-(9) can
be represented in the form of the first-order equation
\[
\partial _\mu \left(\varepsilon ^{\nu,[\nu\mu]}+ \varepsilon ^{[\nu\mu],\nu}
+ \varepsilon ^{\widetilde{\nu},\widetilde{[\nu\mu]}}+ \varepsilon
^{\widetilde{[\nu\mu]},\widetilde{\nu}}\right)_{AB}\Psi _B(x)
\]
\vspace{-7mm}
\begin{equation} \label{13}
\end{equation}
\vspace{-7mm}
\[
+ m\left(\frac{1}{2}\varepsilon ^{[\nu\mu],[\nu\mu]}+ \varepsilon
^{\nu,\widetilde{\nu}}+\varepsilon
^{\widetilde{\nu},\widetilde{\nu}}+ \frac{1}{2}\varepsilon
^{\widetilde{[\nu\mu]},\widetilde{[\nu\mu]}}\right) _{AB}\Psi
_B(x)=0 .
\]
There is a summation over all repeated indices. We define
20-dimensional matrices as follows:
\begin{equation}
\beta_\mu=\beta_\mu^{(1)}+ \widetilde{\beta}_\mu^{(1)},~~~
\beta_\mu^{(1)}=\varepsilon ^{\nu,[\nu\mu]}+ \varepsilon
^{[\nu\mu],\nu},~~~ \widetilde{\beta}_\mu^{(1)}= \varepsilon
^{\widetilde{\nu},\widetilde{[\nu\mu]}}+ \varepsilon
^{\widetilde{[\nu\mu]},\widetilde{\nu}},
 \label{14}
\end{equation}
\begin{equation}
P=\frac{1}{2}\varepsilon ^{[\nu\mu],[\nu\mu]}+ \varepsilon
^{\nu,\widetilde{\nu}}+\varepsilon
^{\widetilde{\nu},\widetilde{\nu}}+ \frac{1}{2}\varepsilon
^{\widetilde{[\nu\mu]},\widetilde{[\nu\mu]}}. \label{15}
\end{equation}
Taking into account Eq.(14),(15), Eq.(13) takes the form of the
first-order relativistic wave equation:
\begin{equation}
\left( \beta _\mu \partial _\mu +mP\right) \Psi (x)=0 . \label{16}
\end{equation}
The presence of the projection operator $P$ in Eq.(16) is
connected with the fact that there is a massless state in the
spectrum \cite{Kruglov}, \cite{Kruglov1}. Thus, we reformulated
the higher derivative equation (2) in the form of the first-order
equation (16). The $P$ is the projection operator, $P^2=P$
\cite{Fedorov} and it is not the Hermitian matrix $P^+\neq P$. The
matrices $\beta_\mu^{(1)}$ and $\widetilde{\beta}_\mu^{(1)}$ are
Hermitian matrices and have non-zero components in 10-dimensional
subspaces $(\mu,[\mu\nu])$,
$(\widetilde{\mu},\widetilde{[\mu\nu]})$, respectively and obey the
Petiau-Duffin-Kemmer algebra \cite{Duffin}, \cite{Kemmer} (see
also \cite{Bogush}, \cite{Kruglov}):
\begin{equation}
\beta _\mu \beta _\nu \beta _\alpha +\beta _\alpha \beta _\nu
\beta _\mu =\delta _{\mu \nu }\beta _\alpha+\delta _{\alpha \nu
}\beta _\mu . \label{17}
\end{equation}
Therefore, the matrix $\beta_\mu$ is the direct sum of two
10-dimensional Petiau-Duffin-Kemmer matrices. The projection
operator $P$ ``connects" two 10-dimensional subspaces
$(\mu,[\mu\nu])$, and $(\widetilde{\mu},\widetilde{[\mu\nu]})$.
Thus, HD Podolsky's electrodynamics equations lead to ``doubling"
the dimension of the Petiau-Duffin-Kemmer algebra representation.

\section{The Lorentz covariance and Hermitianizing matrix}

Let us prove the Lorentz covariance of Eq.(16). The Lorentz group transformations of coordinates are given by
$x_\mu ^{\prime }=L_{\mu \nu }x_\nu ^{\prime } $, where the Lorentz matrix $L=\{L_{\mu \nu }\}$ satisfies the equation
$L_{\mu \alpha }L_{\nu \alpha }=\delta _{\mu \nu } $.
The wave function (11), under the Lorentz coordinates transformations,
becomes
\begin{equation}
\Psi ^{\prime }(x^{\prime })=T\Psi (x) , \label{18}
\end{equation}
where $20\times 20-$matrix $T$ realizes the reducible tensor representation
of the Lorentz group. The first-order wave equation (16) is
transformed into
\begin{equation}
\left( \beta _\mu \partial _\mu ^{\prime }+mP\right) \Psi ^{\prime
}(x^{\prime })=\left( \beta _\mu L_{\mu \nu }\partial _\nu
+mP\right) T\Psi (x)=0 , \label{19}
\end{equation}
where $\partial _\mu ^{\prime }=L_{\mu \nu
}\partial _\nu $. We have the Lorentz covariance of Eq.(16)
if equations
\begin{equation}
\beta_\mu TL_{\mu \nu }=T\beta _\nu,~~~~~~ PT=TP  \label{20}
\end{equation}
hold. The infinitesimal Lorentz matrix is given by the
\begin{equation}
L_{\mu \nu }=\delta _{\mu \nu }+\varepsilon _{\mu \nu } ,
\hspace{0.5in} \varepsilon _{\mu \nu }=-\varepsilon _{\nu \mu } ,
\label{21}
\end{equation}
where $\varepsilon _{\mu \nu }$ are six parameters defining rotations and
boosts. The matrix $T$ at the infinitesimal Lorentz transformations
reads
\begin{equation}
T=1+\frac 12\varepsilon _{\mu \nu }J_{\mu \nu } , \label{22}
\end{equation}
where $J_{\mu \nu }$ are generators of the Lorentz group in $20-$dimensional space. With the aid of
Eq.(21),(22) (using the smallness of parameters $\varepsilon
_{\mu \nu }$), we obtain from Eq.(20)
\begin{equation}
\beta _\mu J_{\alpha \nu }-J_{\alpha \nu }\beta_\mu =\delta
_{\alpha \mu }\beta _\nu -\delta _{\nu \mu }\beta _\alpha, ~~~~PJ_{\alpha \nu }=J_{\alpha \nu }P .
\label{23}
\end{equation}
The Lorentz group generators in the 20-dimensional representation
space are given by
\[
J_{\mu\nu}= \beta_\mu\beta_\nu-\beta_\nu\beta_\mu
 \]
\begin{equation}
=\varepsilon ^{\mu,\nu}-\varepsilon ^{\nu,\mu}+\varepsilon
^{[\lambda\mu],[\lambda\nu]}-\varepsilon
^{[\lambda\nu],[\lambda\mu]}
\label{24}
\end{equation}
\[
+\varepsilon ^{\widetilde{\mu},\widetilde{\nu}}- \varepsilon
^{\widetilde{\nu},\widetilde{\mu}}+ \varepsilon
^{\widetilde{[\lambda\mu]},\widetilde{[\lambda\nu]}}- \varepsilon
^{\widetilde{[\lambda\nu]},\widetilde{[\lambda\mu]}},
\]
and obeys Eq.(23). Thus, we have proved the Lorentz covariance of first-order wave equation (16).
In Appendix A, we generalize equations considered on the case of field equations with the source.
It is easy to verify with the help of Eq.(12) that the generators
(24) obey the usual commutation relations
\begin{equation}
\left[ J_{\mu \nu },J_{\alpha \beta}\right] =\delta _{\nu \alpha
}J_{\mu \beta}+\delta _{\mu \beta }J_{\nu \alpha}-\delta _{\nu
\beta }J_{\mu \alpha}-\delta _{\mu \alpha }J_{\nu \beta}.
\label{25}
\end{equation}
The Hermitianizing matrix
$\eta$ should satisfy the relations \cite{Gel'fand}
\begin{equation}
\eta \beta _m=-\beta _m^{+}\eta^{+} ,\hspace{0.5in}\eta \beta
_4=\beta _4^{+}\eta^{+} \hspace{0.5in}(m=1,2,3) .  \label{26}
\end{equation}
We find
\[
\eta=\varepsilon^{m,m}-\varepsilon^{4,4}+\varepsilon^{[m4],[m4]}-
\frac{1}{2}\varepsilon^{[mn],[mn]}
\]
\vspace{-7mm}
\begin{equation} \label{27}
\end{equation}
\vspace{-7mm}
\[
+\varepsilon^{\widetilde{m},\widetilde{m}} -
\varepsilon^{\widetilde{4},\widetilde{4}} +
\varepsilon^{\widetilde{[m4]},\widetilde{[m4]}} -
\frac{1}{2}\varepsilon^{\widetilde{[mn]},\widetilde{[mn]}} .
\]
The matrix $\eta $ is the Hermitian matrix, $\eta^{+}=\eta$ and
commutes with the projection operator $P$:
\begin{equation}
\eta P=P\eta.
\label{28}
\end{equation}
Consider the ``conjugated" wave function
\begin{equation}
\overline{\Psi }(x)=\Psi ^{+}(x)\eta=
\left(\psi_\mu,-\psi_{\mu\nu},
\widetilde{\psi}_\mu,-\widetilde{\psi}_{\mu\nu}\right), \label{29}
\end{equation}
and $\Psi ^{+}(x)$ is the Hermitian conjugated wave function. We
took into account that for neutral fields, $(\psi_m,\psi_0)$ are
real variables. Thus, the relativistically invariant bilinear form
is $ \overline{\Psi }(x)\Psi (x)=\Psi ^{+}(x)\eta \Psi(x)$. Then,
we obtain from Eq.(16) the ``conjugated" equation
\begin{equation}
\overline{\Psi }(x)\left( \beta _\mu \overleftarrow{\partial} _\mu
-mP^+\right) =0 .
\label{30}
\end{equation}
Formally, one can construct the Lagrangian
\begin{equation}
{\cal L}=-\frac{1}{2}\left[\overline{\Psi }(x)\left(\beta _\mu
\partial _\mu +mP\right) \Psi (x)-\overline{\Psi }(x)\left(\beta _\mu
\overleftarrow{\partial} _\mu -mP^+\right) \Psi (x)\right].
\label{31}
\end{equation}
By varying the action $S=\int d^4x {\cal L}$, corresponding to the
Lagrangian (31), we obtain equations of motion (16), (30). One can
check using Eq.(26),(29) that the Lagrangian ${\cal L}$ is the
real function, ${\cal L}^*={\cal L}$. In addition, for neutral
fields the equality
\begin{equation}
\overline{\Psi }(x)P^+\Psi (x)=\overline{\Psi }(x)P\Psi (x) . \label{32}
\end{equation}
is valid although $P^+\neq P$. If one wants to consider charged fields (not photon fields),
then the electric current density is given by
\begin{equation}
J_\mu (x)=\frac{i}{m^3}\left[\left(\partial_\rho F^\ast_{\rho\nu} \right)\partial_\alpha^2F_{\nu\mu}-\left(\partial_\alpha^2F_{\nu\mu}^\ast\right)
\left(\partial_\rho F_{\rho\nu} \right)\right] , \label{33}
\end{equation}
were the complex conjugation $\ast$ does not act on the metric imaginary unit.
Using equations of motion (2), one can verify that electric current is conserved
$\partial_\mu J_\mu (x)=0$. The electric current density (33) can be cast into the matrix form
\begin{equation}
J_\mu (x)=i\overline{\Psi }(x)\widetilde{\beta}_\mu^{(1)} \Psi(x) . \label{34}
\end{equation}
It follows from Eq.(33) that for the neutral (photon) fields the electric current density
vanishes, $J_\mu (x)=0$, as it should be.

\section{The mass and spin projection operators}

Let us consider solutions to Eq.(16) with definite energy and
momentum. In the momentum space Eq.(16) becomes
\begin{equation}
\Lambda \Psi(p)=0 ,~~~~\Lambda=i\hat{p}+mP,~~~~\hat{p}=\beta_\mu
p_\mu, \label{35}
\end{equation}
where $p_\mu$ is a four-momentum $p_\mu=(\textbf{p},ip_0)$. Let us consider
the massive state, $p^2=-m^2$. For this case
the 20-dimensional matrix $\Lambda$ obeys the equation (see (B5)
in Appendix B)
\begin{equation}
\Lambda\left(\Lambda-m\right)\left(\Lambda-2m\right)
\left(\Lambda^2-m\Lambda-m^2\right)=0
. \label{36}
\end{equation}
From Eq.(36), we find the solution to Eq.(35) in the form of the
matrix
\begin{equation}
 \Pi=N\left(\Lambda-m\right)\left(\Lambda-2m\right)
 \left(\Lambda^2-m\Lambda-m^2\right), \label{37}
\end{equation}
where $N$ is a normalization constant, so that $\Lambda\Pi=0$.
This means the the every column of the matrix $\Pi$ is the
solution to Eq.(35). The requirement that the $\Pi$ is the
projection operator, $\Pi^2=\Pi$, leads to the normalization
constant $N=-1/(2m^4)$ \cite {Fedorov}. The projection operator
(37) extracts solutions to Eq.(35) for definite energy and
momentum corresponding to the massive state.

With the help of the Lorentz group generators (24), we obtain the
spin operator (see \cite{Fedorov}):
\begin{equation}
\sigma _p=-\frac i{2\mid \mathbf{p}\mid }\epsilon
_{abc}p_aJ_{bc}=-\frac i{\mid \mathbf{p}\mid }\epsilon
_{abc}p_a\beta _b\beta _c. \label{38}
\end{equation}
The operator (38) obeys the ``minimal" matrix equation:
\begin{equation}
\sigma _p\left( \sigma _p-1\right) \left( \sigma _p+1\right) =0 .
\label{39}
\end{equation}
In accordance with the general method \cite{Fedorov}, we obtain
the projection operators extracting spin projections $\pm 1$ and
$0$:
\begin{equation}
S_{(\pm 1)}=\frac 12\sigma _p\left( \sigma _p\pm 1\right)
,\hspace{0.5in}S_{(0)}=1-\sigma _p^2,\label{40}
\end{equation}
satisfying the relations: $S_{(\pm 1)}^2=S_{(\pm 1)}$, $S _{(\pm
1)}S_{(0)}=0$, $S_{(0)}^2=S_{(0)}$.

One may check with the help of Eq.(12) that the operators (40)
commute with the mass projection operator (37). As a result, from
Eq.(37),(40), we find projection operators
\begin{equation}
\Delta_{\pm 1} =\Pi S_{(\pm 1)} ,~~~~\Delta _0=\Pi S_{(0)}
\label{41}
\end{equation}
extracting solutions to Eq.(35) for definite energy-momentum, spin
projections $\pm 1$, $0$, for states of particles with the mass
$m$. Eq.(41) defines also the density matrix for pure spin states.
It follows from ``minimal" polynomial equation (B4) that for the massless state,
$p^2=0$, zero eigenvalues of the matrix $\Lambda$ are degenerated, and therefore it is
imposable to construct solutions to Eq.(35) in the form of the projection operator \cite{Fedorov}.

\section{Quantum mechanical Hamiltonian}

Now we obtain the quantum mechanical Hamiltonian
from equations (6)-(9). The Schrodinger form of equations has some
attractive features because non-dynamical components of the wave function
are absent. To find the Schrodinger form of Eq.(6)-(9), we exclude the non-dynamical components.
Eq.(6)-(9) can be cast in the form of two systems
\[
m\psi _{4 m}=\partial_4 \psi_m -\partial _m \psi_4 , ~~~~
m\widetilde{\psi}_{4m}=\partial_4 \widetilde{\psi}_m -\partial_m \widetilde{\psi}_4 ,
\]
\vspace{-8mm}
\begin{equation} \label{42}
\end{equation}
\vspace{-8mm}
\[
\partial_4 \psi_{m4}+\partial_n \psi_{m n} =
-m\widetilde{\psi}_m,~~~~~~\partial_4 \widetilde{\psi}_{m4}+\partial_n \widetilde{\psi}_{m n} =
-m\widetilde{\psi}_m ,
\]
\begin{equation}
m\psi_{mn}=\partial_m \psi_n -\partial_n \psi_m , ~~~~
m\widetilde{\psi}_{mn}=\partial_m \widetilde{\psi}_n -\partial_n \widetilde{\psi}_m, ~~~~m\widetilde{\psi}_4=\partial_m\widetilde{\psi}_{m4}.
\label{43}
\end{equation}
We can to exclude auxiliary (non-dynamical) components $\psi_{mn}$, $\widetilde{\psi}_{mn}$,
$\widetilde{\psi}_4$ from Eq.(43).
However, the $\psi_4$ can not be excluded from Eq.(42).
To introduce the evolution of the $\psi_4$ in time, we use the Lorentz
condition $\partial_m\psi_m+\partial_4\psi_4=0$. After the
replacing the non-dynamical components $\psi_{mn}$, $\widetilde{\psi}_{mn}$,
$\widetilde{\psi}_4$ from Eq.(43) into Eq.(42), we obtain the equations as follows:
\[
i\partial _t \psi _m=m\psi _{m4} -\partial_m \psi _4 ,
\]
\[
i\partial _t \psi _4=\partial_n \psi _n ,
\]
\begin{equation}
i\partial _t \widetilde{\psi}_m=m\widetilde{\psi}_{m4} -\partial_m \widetilde{\psi}_4 ,
 \label{44}
\end{equation}
\[
i\partial _t \psi_{n4}=m\widetilde{\psi}_n + \frac{1}{m}\left(\partial_m\partial
_n \psi _m-\partial_m^2\psi_n\right) ,
\]
\[
i\partial _t \widetilde{\psi}_{n4}=m\widetilde{\psi}_n + \frac{1}{m}\left(\partial_m\partial
_n \widetilde{\psi}_m-\partial_m^2\widetilde{\psi}_n\right) .
\]
Eq.(44) show that 13-components of the wave function $\Psi(x)$ possess the evolution in time. Therefore, we
introduce the 13-component wave function
\begin{equation}
\Phi (x)=\left(
\begin{array}{c}
\psi_\mu (x)\\
\psi_{m4}(x)\\
\widetilde{\psi}_m(x)\\
\widetilde{\psi}_{m4}(x)
\end{array}
\right).\label{45}
\end{equation}
With the help of the elements of the matrix algebra Eq.(12), we
rewrite Eq.(44) in the Schrodinger form
\begin{equation}
i\partial_t\Phi(x)={\cal H}\Phi(x),
\label{46}
\end{equation}
where the Hamiltonian is given by
\[
{\cal
H}=m\left(\varepsilon^{n,[n4]}+\varepsilon^{\widetilde{n},\widetilde{[n4]}}+
\varepsilon^{[n4],\widetilde{n}}+\varepsilon^{\widetilde{[n4]},\widetilde{n}}\right)
+\left(\varepsilon^{4,m}-\varepsilon^{m,4}\right)\partial_m
\]
\vspace{-8mm}
\begin{equation} \label{47}
\end{equation}
\vspace{-8mm}
\[
+ \frac{1}{m}\biggl[\left(\varepsilon^{[m4],n}+\varepsilon^{\widetilde{[m4]},\widetilde{n}}
-\varepsilon^{\widetilde{n},\widetilde{[m4]}}\right)\partial_m\partial_n\
-\left(\varepsilon^{[m4],m}+\varepsilon^{\widetilde{[m4]},\widetilde{m}}\right)\partial_n^2\biggr].
\]
From the minimal equation (C6), obtained in Appendix C, we find
the projection operators extracting states with positive
and negative energies for the massless states ($p^2=0$)
\begin{equation}
\Sigma^0_{\pm}=\pm\frac{{\left({\cal H}\pm
|\textbf{p}|\right)\cal H}^2\left({\cal H}^2-\textbf{p}^2-m^2\right)\left({\cal H}^2-2
\textbf{p}^2-m^2\right)}{2|\textbf{p}|^3m^2\left(\textbf{p}^2+m^2\right)},\label{48}
\end{equation}
and massive states ($p^2=-m^2$)
\begin{equation}
\Sigma_{\pm}=\mp\frac{\left({\cal H}\pm
p_0\right){\cal H}^2\left({\cal H}^2-\textbf{p}^2\right)\left({\cal H}^2-
\textbf{p}^2-p_0^2\right)}{2p_0^3m^2\textbf{p}^2}.\label{49}
\end{equation}
Projection operators (48),(49) obey equations as follow:
\[
\left(\Sigma_{\pm}^0\right)^2=\Sigma^0_{\pm},~~~~{\cal H}\Sigma_{\pm}^0=\pm p_0\Sigma^0_{\pm}  ~~~~~~~~\left(p_0=|\textbf{p}|\right),
\]
\vspace{-8mm}
\begin{equation} \label{50}
\end{equation}
\vspace{-8mm}
\[
\left(\Sigma_{\pm}\right)^2=\Sigma_{\pm},~~~~{\cal H}\Sigma_{\pm}=\pm p_0\Sigma_{\pm}~~~~~~~~~\left(p_0=\sqrt{|\textbf{p}|^2+m^2}\right).
\]
Projection operators (48),(49) can be used to construct physical states in 13-dimensional space of wave functions
(45).

\section{Conclusion}

We have formulated Podolsky's generalized electrodynamics equation
with higher derivatives in the form of the 20-component
first-order relativistic wave equation. This equation describes
vector particles possessing the physical massless state and the
massive state that is a ghost. To have the consistent theory the mass of the vector state should be very large.
One can speculate that the massive vector
particles can be described in the gauge-invariant manner by this theory. To have the massive state
to be the physical state, we have to use the reverse-sign in the Lagrangian. Then the Hamiltonian also changes the sign. In this case, however, the massless state becomes the ghost and the question arises: how to get rid of it?
Therefore the description of massive particles by Podolsky's generalized electrodynamics is questionable.
The relativistically invariant bilinear form, and the Lagrangian were obtained, and these allow us to use
the advantages of the formulation of relativistic wave equations.
The density matrix obtained can be used for quantum electrodynamics calculations in the first-order formalism.
It should be noted that the Petiau-Duffin-Kemmer form of equations was used
in quantum chromodynamics \cite{Gribov}, i.e. in non-Abelian theory.

The $13\times 13$-matrix Schrodinger form of the equation is derived, and the Hamiltonian is obtained.
We found projection operators extracting the physical eigenvalues of the Hamiltonian.
The Schrodinger picture has some advantages by considering field interactions.

\vspace{7mm}
\textbf{Appendix A}
\vspace{7mm}

Let us consider the field equation (2) with the source of electromagnetic fields $-$ the charge current
density:
$$
\left(\partial_\alpha^2 -m^2 \right)\partial_\mu F_{\nu\mu}=-m^2\widetilde{j}_\nu.~~~~~~~~~~~~~~~~~~~~~~(A1)
$$
We have introduced the current $\widetilde{j}_\nu$ with the same dimension as in classical electrodynamics.
The first-order equations (6),(7),(9) reman the same but Eq.(8) is replaced by
$$
\partial_\mu\widetilde{\psi}_{\nu\mu} + m \widetilde{\psi}_\nu=\widetilde{j}_\nu(x).~~~~~~~~~~~~~~~~~~~~~~~~~~~~~~(A2)
$$
Then Eq.(16) becomes
$$
\left( \beta _\mu \partial _\mu +mP\right) \Psi (x)=P_0 j(x),~~~~~~~~~~~~~~~~~~~~~~~(A3)
$$
where
$$
P_0=\varepsilon^{\widetilde{\mu},\widetilde{\mu}},~~~~j(x)=\left(
\begin{array}{c}
j_\mu(x)\\
j_{\mu\nu}(x)\\
\widetilde{j}_\mu(x)\\
\widetilde{j}_{\mu\nu}(x)
\end{array}
\right), ~~~~~~~~~~~~~~~~~~~(A4)
$$
and $P_0$ is the projection operator, $P_0^2=P_0$, $P_0^+=P_0$. The projection operator $P_0$ extracts only the current $\widetilde{j}_\mu$. Therefore, the currents $j_\mu(x)$, $j_{\mu\nu}(x)$, $\widetilde{j}_{\mu\nu}(x)$ do not present in the theory and can be put zero. At the Lorentz transformations, $j'(x)=Tj(x)$, and the Lorentz covariance of Eq.(A3) follows from Eq.(20),(23) and
$$
P_0T=TP_0, ~~~~~~~~~P_0J_{\mu\nu}=J_{\mu\nu}P_0.~~~~~~~~~~~~~~~~~~~(A5)
$$
The Hermitianizing matrix $\eta$ (27) commutes with $P_0$, $\eta P_0=P_0\eta$. Then Eq.(30) is replaced by
$$
\overline{\Psi }(x)\left( \beta _\mu \overleftarrow{\partial} _\mu
-mP^+\right) =\overline{j}(x)P_0 ,~~~~~~~~~~~~~~~~~~~~(A6)
$$
where $\overline{j}(x)=\left(j_\mu(x),-j_{\mu\nu}(x),\widetilde{j}_\mu(x),-\widetilde{j}_{\mu\nu}(x)\right)$.
We obtain the classical limit at $m\rightarrow\infty$ ($a\rightarrow 0$) for Maxwellian electrodynamics from Eq.(A1):
$$
\partial_\mu F_{\nu\mu}=\widetilde{j}_\nu.~~~~~~~~~~~~~~~~~~~~~~~~~~~~~~~~~~~~~~~(A7)
$$
Thus, Eq.(A7) is the standard Maxwell equation with the source term.

\vspace{7mm}
\textbf{Appendix B}
\vspace{7mm}

With the help of Eq.(12), we obtain products of matrices entering
Eq.(35):
$$
\hat{p}^3=p^2\hat{p},~~~~\hat{p}P+P\hat{p}=P\hat{p}P+\hat{p},~~~~\hat{p}^2P=P\hat{p}^2,
~~~~~~~~~~~~~~~~~~~~~~(B1)
$$
$$
\hat{p}P\hat{p}^2=p^2\hat{p}P,~~~~\hat{p}P\hat{p}(1-P)=\hat{p}^2(1-P),~~~~(1-P)\hat{p}^2P=0.~~~~~~(B2)
$$
Using Eq.(14), (B1), (B2), one finds
$$
\Lambda\left(\Lambda-m\right)=imP\hat{p}P-\hat{p}^2,~~~~
$$
$$
\Lambda\left(\Lambda-m\right)\left[\Lambda\left(\Lambda-m\right)^2+2p^2\left(\Lambda-m\right)
+mp^2\right]=-ip^4\hat{p}-mp^2\hat{p}^2P,~~~(B3)
$$
$$
\Lambda\left(\Lambda-m\right)\left[\Lambda\left(\Lambda-m\right)-m\left(\Lambda-m\right)
+2p^2\right]=imp^2\hat{p}-p^2\hat{p}^2-m^2\hat{p}^2\left(1-P\right).
$$
From Eq.(B1)-(B3), we obtain ``minimal" polynomials of the matrix
$\Lambda$ for two states:
$$
\Lambda^2\left(\Lambda-m\right)^3=0,~~~~p^2=0,~~~~~~~~~~~~~~~~~~~~~~~~~~~~~~~~~~~~~~~~~~~~~~~~~~~~(B4)
$$
$$
\Lambda\left(\Lambda-m\right)\left(\Lambda-2m\right)
\left(\Lambda^2-m\Lambda-m^2\right)=0 ,~~~~p^2=-m^2.~~~~~~~~~~~~~~~ (B5)
$$
It should be noted that zero eigenvalues of the matrix $\Lambda$ for the massless state are degenerated.

\vspace{7mm}
\textbf{Appendix C}
\vspace{7mm}

From Eq.(47), we obtain the Hamiltonian in the momentum space
$$
{\cal H}=m\left(\varepsilon^{n,[n4]}+\varepsilon^{\widetilde{n},\widetilde{[n4]}}+
\varepsilon^{[n4],\widetilde{n}}+\varepsilon^{\widetilde{[n4]},\widetilde{n}}\right)
+ip_m\left(\varepsilon^{4,m}-\varepsilon^{m,4}\right)
$$
$$
+ \frac{1}{m}\biggl[\left(\varepsilon^{[m4],m}+\varepsilon^{\widetilde{[m4]},\widetilde{m}}\right)\textbf{p}^2
-\left(\varepsilon^{[m4],n}+\varepsilon^{\widetilde{[m4]},\widetilde{n}}
-\varepsilon^{\widetilde{n},\widetilde{[m4]}}\right)p_m p_n\biggr].~~~~(C1)
$$
Using Eq.(12), one finds
$$
{\cal H}^2-\textbf{p}^2=m^2\left(\varepsilon^{[n4],\widetilde{[n4]}}
+\varepsilon^{\widetilde{[n4]},\widetilde{[n4]}}
+\varepsilon^{\widetilde{n},\widetilde{n}}+\varepsilon^{n,\widetilde{n}}\right)
$$
$$
+imp_n\varepsilon^{4,[n4]}-p_m p_n\left(\varepsilon^{[n4],[m4]}- \varepsilon^{[n4],\widetilde{[m4]}}\right),
~~~~~~~~~~~~~~~~~~~~~~~~~~~~~(C2)
$$
$$
{\cal H}^2-\textbf{p}^2-m^2=m^2\left(\varepsilon^{[n4],\widetilde{[n4]}}
-\varepsilon^{[n4],[n4]}
+\varepsilon^{n,\widetilde{n}}-\varepsilon^{\mu,\mu}\right)
$$
$$
+imp_n\varepsilon^{4,[n4]}-p_m p_n\left(\varepsilon^{[n4],[m4]}- \varepsilon^{[n4],\widetilde{[m4]}}\right).
~~~~~~~~~~~~~~~~~~~~~~~~~~~~~~(C3)
$$
Multiplying Eq.(C2) and Eq.(C3), we obtain
$$
\left({\cal H}^2-\textbf{p}^2\right)\left({\cal H}^2-\textbf{p}^2-m^2\right)=im\left(m^2+\textbf{p}^2\right)
p_m\left(\varepsilon^{4,\widetilde{[m4]}}-\varepsilon^{4,[m4]}\right)
$$
$$
+
\left(m^2+\textbf{p}^2\right)p_mp_n\left(\varepsilon^{[n4],[m4]}-\varepsilon^{[n4],\widetilde{[m4]}}\right).
~~~~~~~~~~~~~~~~~~~~~~~~~~~~~~~~(C4)
$$
Squiring Eq.(C4), one finds
$$
\left({\cal H}^2-\textbf{p}^2\right)^2\left({\cal H}^2-\textbf{p}^2-m^2\right)^2
$$
$$
= \textbf{p}^2\left(m^2+\textbf{p}^2\right)\left({\cal H}^2-\textbf{p}^2\right)\left({\cal H}^2-\textbf{p}^2-m^2\right).
~~~~~~~~~~~~~~~~~~~~~~~~~~(C5)
$$
From Eq.(C5), we obtain the ``minimal" polynomial of the Hamiltonian
$$
{\cal H}^2\left({\cal H}^2-\textbf{p}^2\right)\left({\cal H}^2-\textbf{p}^2-m^2\right)\left({\cal H}^2-2\textbf{p}^2-m^2\right)=0.
~~~~~~~~~~~~~~(C6)
$$
Eigenvalues of the Hamiltonian squared read from Eq.(C6): $p_0^2=0$, $p_0^2=\textbf{p}^2$,
$p_0^2=\textbf{p}^2+m^2$, $p_0^2=2\textbf{p}^2+m^2$. Thus, there are two physical eigenvalues,
$p_0^2=\textbf{p}^2$, $p_0^2=\textbf{p}^2+m^2$, corresponding to massless and massive states of the field, and two
nonphysical eigenvalues. Eq.(C6) can be used to find projection operators extracting physical states in the Schrodinger
picture.

\end{document}